# Nanoscale perspective on the stress-corrosion cracking behavior of a peak-aged 7XXX-Al alloy

*Martí López Freixes[1], Lionel Peguet[2], Timothy Warner[2], Baptiste Gault[1,3]\**

[1] *Max-Planck-Institut für Eisenforschung GmbH, Max-Planck-Str. 1, 40237 Düsseldorf, Germany*

[2] *C-TEC, Constellium Technology Center, Parc Economique Centr'alp, CS 10027, Voreppe, 38341 cedex, France*

[3] *Department of Materials, Royal School of Mines, Imperial College London, Prince Consort Road, London SW7 2BP, UK*

*\* corresponding authors. E-mail addresses: b.gault@mpie.de*

**Abstract**

High strength 7xxx Al-alloys are currently commonly used in aerospace and are expected to be increasingly employed in the automotive sector for weight reduction purposes. These alloys can however be sensitive to stress-corrosion cracking (SCC) depending on temper and loading conditions. Both the alloy's grain structure and composition are believed to play a key role in determining sensitivity to SCC. Here, we study at the nanometer scale the evolution of the microstructure near stress corrosion cracks on two different model variants of the 7140 aluminum alloy. We performed double cantilever beam (DCB) crack growth tests in hot (70°C) humid air, on samples extracted at quarter-thickness (T/4) and mid-thickness (T/2) and heat treated to a non-industrial, SCC sensitive T6 condition. The sample at T/4 shows a lower $K_{ISCC}$ along with flatter grains and a higher solute content, whereas both samples exhibit similar crack growth rates at higher stress intensities. We report on precipitate dissolution and matrix solute enrichment near the crack tips, with the T/4 position presenting the higher increase in solute levels. The near grain boundary microstructure ahead of the crack is modified, with evidence of precipitate dissolution and transport of solutes towards the stress-corrosion crack tip. These results agree with a recent report on another 7xxx Al-alloy after SCC in Cl-solution, supporting the possibility that these mechanisms are generally occurring. We relate our findings with the measured SCC behavior and provide an array of possible mechanisms that could be widely applicable in SCC of high strength Al-alloys.

# 1 Introduction

The high-strength 7xxx series of Al-alloys are currently widely used in aerospace and increasingly in the automotive industry [1,2]. They are expected to play a key role in reducing $CO_2$ emissions in the future as lighter, high-strength alloys are implemented into vehicles to reduce their weight and fuel consumption. However, these alloys can be susceptible to stress-corrosion cracking (SCC) as a function of temper and exposure conditions through mechanisms that remain not fully determined [3–5], making it challenging to define efficient counteracting strategies while maintaining an appropriate property balance.

SCC in 7xxx Al-alloys can occur in a variety of environments under static loading, such as Cl solutions [5,6], distilled water [4] and also in humid air [4,6,7]. It is widely believed that SCC cracks grow via a hydrogen mediated process [4,5,8]. As a consequence of oxidation at the crack tip, atomic H is produced, absorbed in the alloy and is expected to subsequently lead to embrittlement through one or more of the processes found under the umbrella term of hydrogen embrittlement (HE). A myriad of mechanisms has indeed been proposed in the scientific literature [9,10], with hydrogen enhanced decohesion (HEDE) [11], hydrogen enhanced localized plasticity (HELP) [12] and adsorption induced localized plasticity (AIDE) [13] being the most prevalent in SCC of high-strength alloys.

In 7xxx Al-alloys strength is achieved through precipitation of the η-phase (Mg(Zn, Cu, Al)$_2$) in the grain interior. At grain boundaries (GBs), intermetallic particles, dispersoids and also η-phase precipitates are present, creating complex microstructures [14]. SCC cracks are almost exclusively intergranular, which is attributed to the electrochemical reactivity of the GB region compared to the matrix [3–5,15] along with the propensity of H to segregate to GBs [16,17], although the relatively soft precipitate-free zone may also play a role [18]. The microstructure and microchemistry of the alloys can be tailored to improve resistance against SCC, with Cu additions and overaging often the preferred routes [15,19].

In thick-plate products, most material characteristics are not homogeneous through the thickness, and in particular HE and SCC resistance [18]. Susceptibility to SCC tends to be lower towards the centerline of the plate compared to the outer parts closer to the surface, although this effect seems less pronounced in higher Cu-containing 7xxx Al-alloys [15,18–20]. The observed improvement against SCC has been mainly attributed to coarser, more widely spaced GB precipitates containing more Cu, associated to a lower quench rate at the center of the plate. In general, a slower quench reduces the stage II plateau crack growth rate and increases the critical stress intensity factor ($K_{ISCC}$) [18,19]. A significant effect of grain structure has also been reported, with flatter grains perpendicular to the loading direction increasing SCC sensitivity [7,21]. Differences in crystallographic texture through the plate appear to have a minor effect on SCC performance [18], although studies on this matter are scarce.

Variation in chemical composition across the thickness also develops in thick-plate products. The direct chill (DC) casting process used for these Al-alloys causes irrecoverable macro-segregations in the final plate, with the centerline (T/2) presenting a slight solute depletion and quarter-thickness (T/4) displaying a slight increase [22]. This pattern is shared by the main alloying elements of the 7xxx series: Zn, Mg and Cu, which have a eutectic reaction with Al. Other minor alloying elements like Ti and Cr, have a peritectic reaction with Al, and therefore exhibit the opposite segregation pattern. Macro-segregations along with the different grain structures found at different thickness positions can lead to slightly different microstructures across the plate, especially as the whole product will undergo the same heat treatment. The microstructure-composition variation across the plate can then alter its sensitivity to SCC.

Here, we aim to study the microstructural evolutions near stress-corrosion cracks in 7140 in a T6 state in hot humid conditions and build upon the observations made in another 7xxx Al-alloy on Cl-containing solution [16]. We performed double cantilever beam (DCB) crack growth tests on two different non-industrially-relevant microstructure-composition variants of 7140-T6, extracted from mid-thickness (T/2) and quarter-thickness (T/4). To avoid the confounding effect of the quench rate, the DCB specimens were solution heat treated and quenched after extraction from the plate. These were subsequently aged to peak strength to ensure sensitivity to SCC and maximize the difference in response between the two cases. We characterize the microstructures at T/2 and T/4 by using scanning electron microscopy (SEM) and atom probe tomography (APT), with the T/4 position having the higher solute fraction but its matrix and GB precipitates being lower in Zn, Mg and Cu. The grain structure is flatter at T/4 whereas T/2 presents the most pronounced texture. We provide evidence of η-phase precipitate dissolution and matrix solute enrichment near the crack tip, with the solute-rich T/4 sample exhibiting twice as much enrichment as that in T/2. In addition, presumably due to dislocation activity, the near GB microstructure ahead of the stress-corrosion crack is also modified. We relate our findings with the measured dissimilar SCC behavior and provide a mechanistic explanation for the compositional alterations observed ahead of the crack and how these may influence SCC susceptibility along with other contributing factors. Advancing the understanding of these possible mechanisms can later help guide alloy design to counteract these degradation processes.

## 2 Experimental

SEM microscopy of the fracture surfaces was performed in a Zeiss Merlin operated at 2kV and 1 nA. Electron backscatter diffraction (EBSD) analysis was performed in a Zeiss Sigma 500 operated at 15 kV and 7.4 nA fitted with an EDAX/TSL camera. The scan step size was 1.5 µm and TSL OIM Analysis 7.0 was used to analyse the data.

DCB specimens were machined in the S-L orientation from the mid-thickness (T/2) and quarter-thickness (T/4) of 124 mm 7140-T7451 plate supplied by Constellium (Table 1). These were then solution heat treated according to the conditions defined for 7140 in AMS 2772G (between 471°C and 482°C) [23] and water quenched. The samples were subsequently aged following a two-step treatment, with a final temperature of 154°C to obtain a state close to peak aged (designated hereafter 7140-T6). This temper is not industrially available, but selected to ensure sensitivity to SCC. DCB crack growth tests were carried out according to ASTM G168-17 [24] at 70°C and 85%RH. Two tests were performed for the T/4 position and 1 test for T/2. The specimens were pre-cracked by fatigue using a sinusoidal cycle at 30 Hz and R = 0,1. The number of cycles required to achieve the target length was about 300,000 for the samples from T/4 and 200,000 for the sample from T/2.

Needle-shaped APT specimens were prepared using an FEI Helios Xe-Plasma focused ion beam (PFIB) following the preparation procedure outlined in ref. [25]. APT analyses were performed on a Cameca Instrument Inc. Local Electrode Atom Probe (LEAP) 5000 XR (fitted with a reflectron) using the voltage-pulsing mode. The oxide containing specimens were run at a base temperature of 80K, with a 15% pulse fraction at a rate of 100 kHz and with 5 ions detected per 1000 pulses. The metallic specimens were acquired at 50K, 20% pulse fraction at 200 kHz and 0.5% detection rate. The raw data was reconstructed and analyzed using AP Suite 6.1 following a crystallography-based calibration method [26].

*Table 1. Nominal compositions for AA7140 in at.% calculated from wt.% nominals*

| Al | Zn | Mg | Cu | Si | Fe | Zr | Ti | Mn | Cr |
|---|---|---|---|---|---|---|---|---|---|
| Bal. | 2.71 | 2.17 | 0.77 | 0.05 | 0.03 | 0.03 | 0.03 | 0.01 | 0.01 |

# 3 Results

## 3.1 Plate microstructure and microchemistry

We performed 2.1 x 0.95 mm EBSD scans to characterize the grain structure around the stress-corrosion cracks in the ST-L orientation at each plate position and determine its possible influence on crack propagation (Supplementary Figure 1). Grain size and recrystallized fraction are similar in both samples but the T/4 position presents flatter grains, as shown by an increased aspect ratio (Supplementary Table 1). Supplementary Figure 2 presents {111} pole figures from both samples, with the T/2 position presenting the stronger texture (Supplementary Table 2).

*Table 2. Reference composition in at. % in the grain interior (matrix + precipitates) near the GB measured by APT. For each plate position, the measurements are the average of 3 datasets, of which 1 was GB containing. The error values correspond to the standard deviation. Supplementary Figure 3 shows the precise location of the measurements within a GB containing dataset.*

|     | Al | Zn | Mg | Cu |
|-----|----|----|----|-----|
| T/2 | 97.42 ± 0.68 | 1.28 ± 0.33 | 0.92 ± 0.15 | 0.37 ± 0.01 |
| T/4 | 96.91 ± 0.84 | 1.53 ± 0.58 | 1.13 ± 0.32 | 0.43 ± 0.08 |

APT measurements of the bulk composition in the grain interior at both plate positions are shown in Table 2. To avoid potential issues with microstructure evolution after long term exposure to 70°C, the analysis was performed on the same samples, near high-angle GBs in regions far from the crack. The total measured solute content in the grain interiors at the T/4 position is higher than at T/2 consistent with the macro-segregation occurring during the DC casting of Al [22]. Figure 1a shows the reconstruction of a typical APT dataset containing a GB, along with the definitions of where the composition measurements of the matrix, η-phase precipitates and GB were made, as summarized in Figure 1b–d. Note that these definitions are used across the entire article and that we refer to η/η' as η-phase precipitates. The values reported are obtained from averaging over multiple datasets for both plate positions.

The difference in the compositions of the grain interior between plate positions (Table 2) modifies the state of precipitation during the subsequent ageing heat treatment. The higher solute content at T/4 increases the degree of supersaturation of the solid solution after quenching, facilitating precipitate nucleation and allowing for faster growth at equivalent heat-treatments [27]. This is confirmed by the lower matrix (Figure 1b) and GB solute (Figure 1c) levels despite the higher total solute content at T/4, characteristic of a later ageing stage [28]. Although we measured three different high-angle GBs, the relatively high fluctuations in the GB solute measurements (Figure 1c) should be noted. Keeping in mind the limited statistics from our APT investigation, η-phase precipitates at the GB appear slightly lower in Zn, Mg and Cu at T/4, as shown in Figure 1d. In the grain interiors, η-phase precipitates follow a linear distribution of possible Zn and Mg values (Figure 1e). The matrix η-phase precipitates at T/2 are slightly higher in Zn and Mg (Figure 1e) but markedly lower in Cu (Supplementary Figure 4).

In the grain interiors imaged in the datasets used to extract the data reported in Figure 1, we determined the number density using precipitates identified through Mg iso-surfaces. This approach may seem simplistic, but attempts at determining number density and volume fraction of precipitates in the grain interior by means of cluster analysis algorithms were deemed too dependent on the input parameters to reliably compare both plate positions. Multiple attempts have demonstrated the complexity of the process, without managing to define a heuristic that can be applied across datasets and materials systems [29,30]. The calculated matrix η-phase precipitate number density sits at 2.82*10e-4 nm-3 for T/2 and 3.38*10e-4 nm-3 for T/4, in line with measurements from the literature performed on a similar alloy [28]. We can then assume a higher precipitate volume fraction at T/4 due to its higher bulk solute content in the grain interior (Table 2), lower matrix solute content (Figure 1b) and a number density slightly higher than that at T/2.

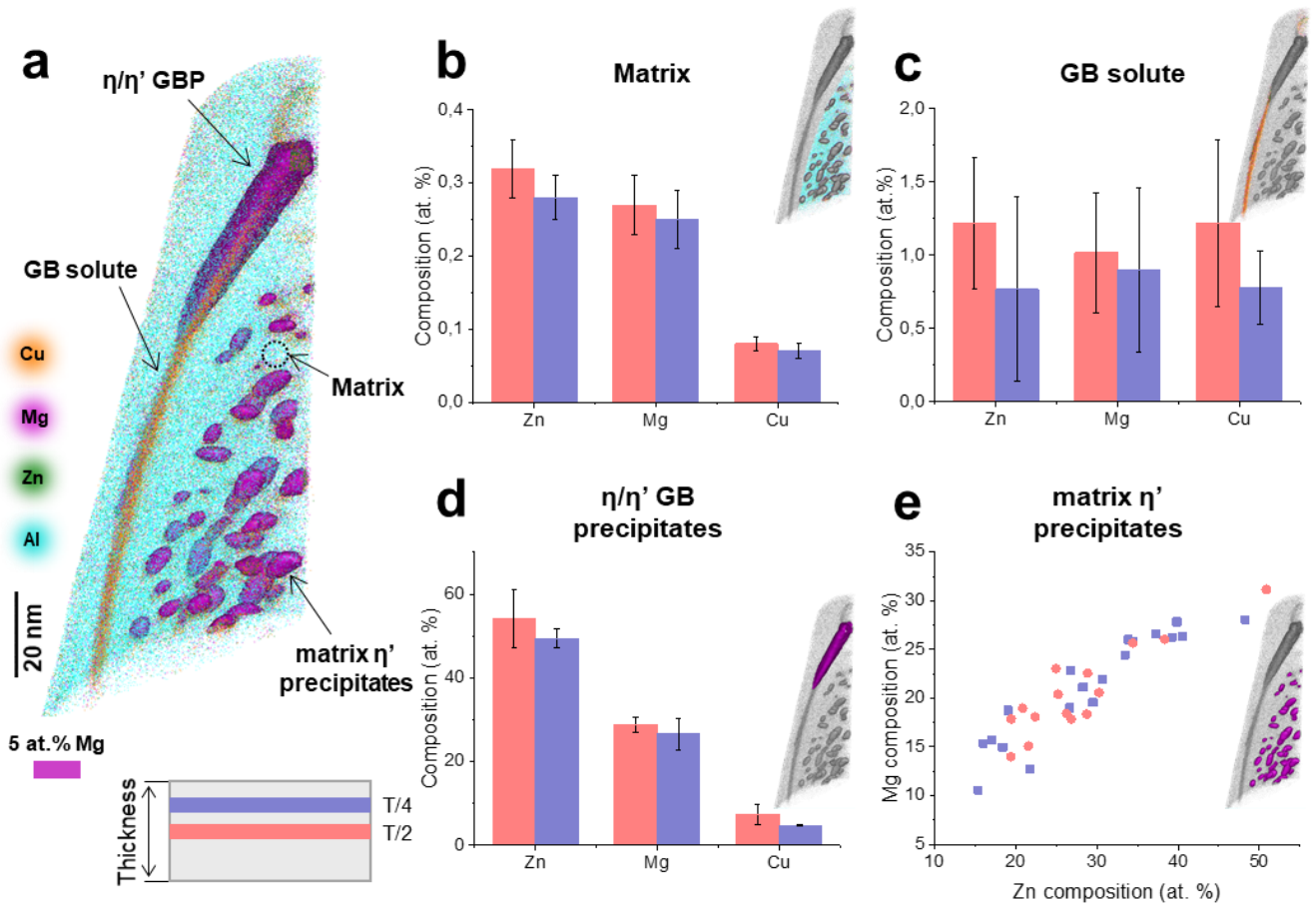

*Figure 1. APT characterization of the microstructure at the two plate positions. a) APT reconstruction of a GB containing reference dataset from the T/2 position showing the microstructural features that are characterized. b) Matrix composition at the two plate positions. The values shown are the average of 3 different datasets from different grains. The error values correspond to the standard deviation between the measurements. c) GB solute levels measured at both plate positions. The values shown are the average of 3 different datasets, taken at different high-angle GBs. The error values correspond to the standard deviation of the measurements. d) Composition of GB precipitates at both plate positions. The measurements shown are from one precipitate per plate position. The error bars correspond to the standard deviation within the precipitate. e) Compositions of the matrix η-phase precipitates within the grain interior at both plate positions. 2 datasets for each plate position are shown and each data point corresponds to an individual matrix η-phase precipitate.*

### 3.2 DCB crack growth rate test

The results of the DCB crack growth rate test performed at 70°C and 85% RH on samples from the AA7140-T6 from T/2 and T/4 are shown in Figure 2a. Both plate positions display identical stage II plateau crack growth rates at ~3e-8 m/s, similar to values reported in the literature [31]. However, they show a dissimilar behavior in region I. The $K_{ISCC}$ for the DCB sample from T/4 is 6 MPa√m whereas the specimen at T/2 does not show crack growth until 10 MPa√m. The plot shows data from 2 tests for the T/4 position and 1 test for T/2. Figure 2b-c shows EBSD scans of the crack tip region in the ST-L plane, showing similar crack paths. The stress-corrosion crack followed primarily high-angle GBs in both samples, with the average misorientation of the cracked boundaries in region I loading conditions being 31° for T/2 and 43° for T/4.

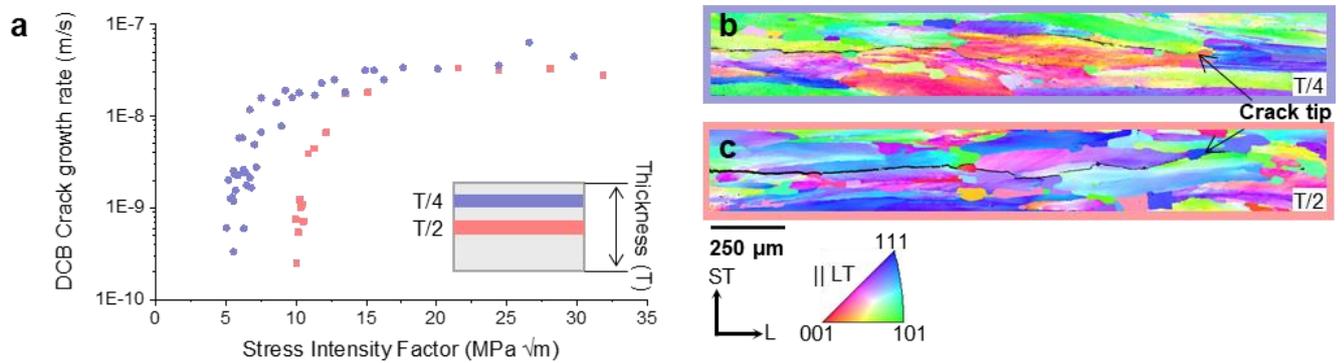

*Figure 2. Crack growth behaviour. a) Crack growth rate double cantilever beam tests for the two plate positions. The results of 2 tests are plotted for T/4 and 1 test for T/2. EBSD map of the final crack tip in the ST-L plate orientation for the b) T/4 position. c) T/2 position*

### 3.3 Fracture surface characterization

The fracture surface observed by SEM shows all features associated with an intergranular mechanism, as expected for SCC in 7xxx Al-alloys and shown in Figure 3. It is mostly covered with a hydroxide of approx. 1µm in thickness, as determined through a PFIB liftout from the fracture surface (Supplementary Figure 5). This suggests liquid water was condensed within the crack during the test, as the oxide thickness after exposure to 70°C and 85%RH is only on the order of tens of nm for 7xxx Al-alloys [32].

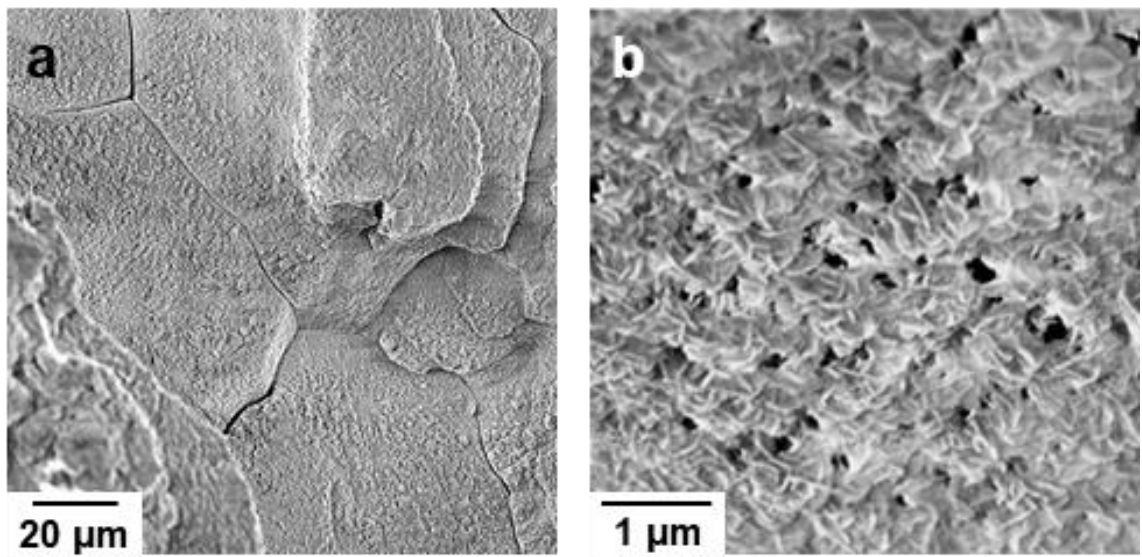

*Figure 3. Characterization of the fracture surface. a) overview of the intergranular fracture surface. b) close-up view of the hydroxide formed.*

### 3.4 Crack tip microchemistry

For both plate positions, we performed APT analyses as near as possible to the stress-corrosion crack tip (Figure 4). The vicinity of the crack is revealed by the presence of an O-rich region highlighted through a set of dark purple iso-surfaces (Figure 4a). Figure 4b–c show composition profiles taken along the matrix-oxide interface for both samples. The oxide is hydrated and enriched in Zn and Mg, and the matrix adjacent to the oxide is also markedly more concentrated in solute compared to the reference measurements shown in Figure 1b.

The composition profiles for both plate positions exhibit only small differences. The average enrichment in Zn, Mg and Cu measured in the matrix adjacent to the oxide (Figure 4b-c) at T/4 is notably higher than at T/2, both in relative and absolute terms, as readily visible in Figure 5. For example, 3.6 at. % Zn is measured at T/4, which is nearly twice the 1.9 at. % Zn at T/2. Zn and Cu follow tendencies consistent with the expected macro-segregation, with T/4 being richer than the alloy's nominal composition, plotted by a dashed line, and T/2 being leaner. Nonetheless, the respective compositional differences are higher than the values calculated with the segregation coefficients for each element [22]. Mg composition within the matrix adjacent to the crack is lower than the nominal in both plate positions, likely due to its preferential oxidation and subsequent loss to the oxide. In addition, matrix η-phase precipitates near the crack tip are lower in Zn, Mg (Figure 6b) and also Cu (Supplementary Figure 6).

Within the oxide layer, the measured solute concentration is higher in Mg, Zn and Cu at T/4, consistent with the higher solute content at this position. The O and H contents are also higher at T/4, with O levels reaching 15 at. % compared to 10 at. % at the T/2 position. These observations confirm the dissolution of precipitates, matrix solute enrichment and oxide characteristics measured near a stress-corrosion crack in NaCl in another 7xxx Al-alloy [16], with maybe unexpected similarity to SCC in hot humid air.

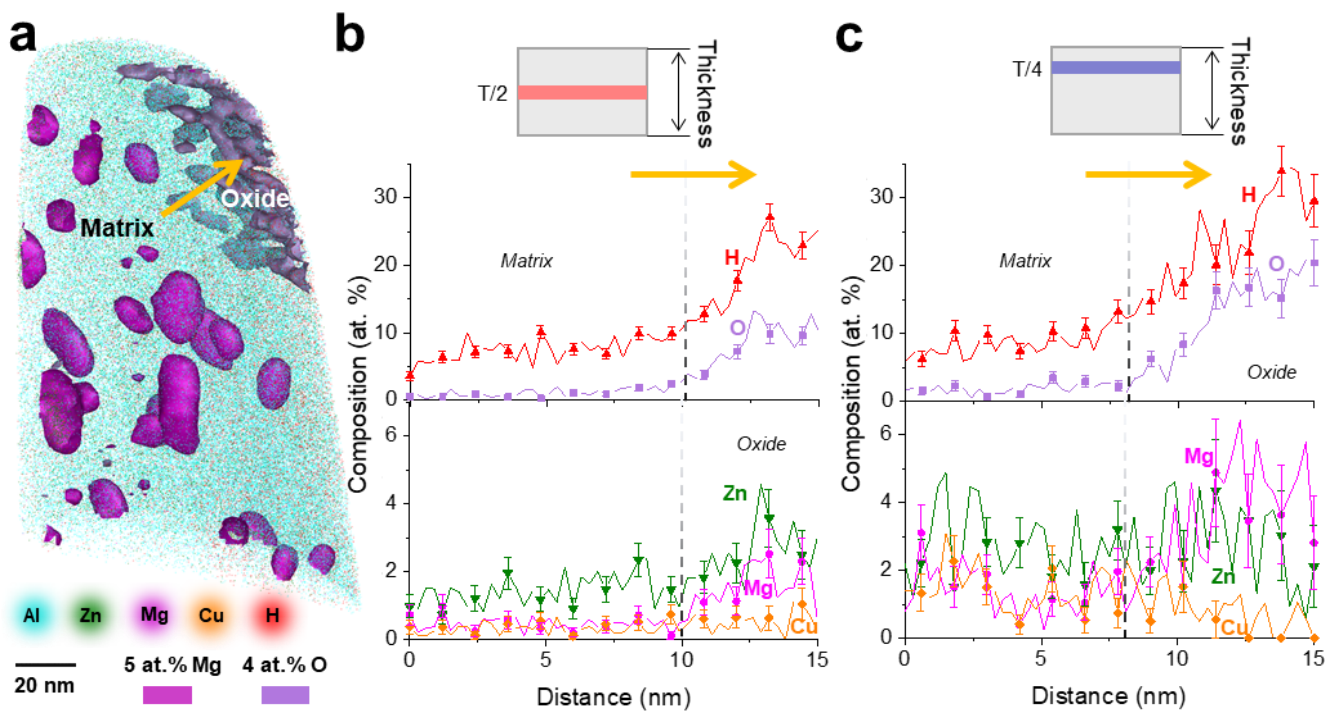

*Figure 4. APT characterization of the crack tips. a) APT reconstruction of a crack tip containing a stress-corrosion crack, displayed through an O iso-composition surface. The dataset shown is from the T/2 position. b) Composition profiles taken along the matrix-oxide interface for the T/2 crack tip, measured within a 10 nm (ø) cylinder. c) Composition profiles taken along the matrix-oxide interface for the T/4 crack tip, measured within a 10 nm (ø) cylinder. The error bars correspond to the standard deviation within each of the bins in the composition profiles shown*

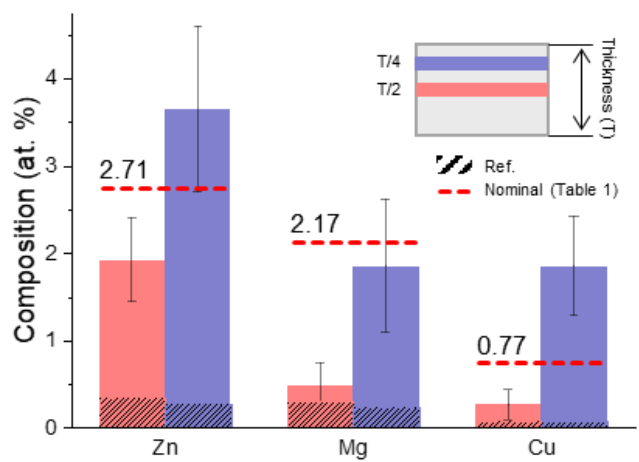

*Figure 5. Composition of the matrix adjacent to the oxide for both plate positions. The values shown are an average of two measurements for T/2 and one for T/4. The plot shows an average of the composition in the matrix adjacent to the oxide, shown in Figure 4. The error values correspond to the standard deviation within the matrix adjacent to the oxide for T/4 and between the two measurements for T/2. The nominal composition of the alloy in at. % (Table 1) is shown by a dashed line and the reference matrix compositions (Figure 1) by diagonal line patterned bars. The error values correspond to the standard deviation within the matrix adjacent to the oxide for T/4 and between the two measurements for T/2.*

### 3.5 Microstructural evolution ahead of the crack

The measured increase in solute levels near the crack tip motivated an analysis of the GB region ahead of the crack. Four APT measurements were performed on the T/2 sample spaced by approx. 2μm, as illustrated in Figure 6a. The compositional evolution of the matrix and matrix η-phase precipitates is plotted in Figure 6b–c as a function of the distance to the crack tip. It should be noted that all datasets except the one at 8 μm contained a section of the same GB that laid ahead of the main intergranular stress-corrosion crack.

The composition of the η-phase precipitates contained in the matrix ahead of the crack, is modified. The composition measurements are equivalent to that shown in Figure 1e. At 2μm, some matrix η-phase precipitates are lower in Zn, Mg (Figure 6b) and higher in Cu (Supplementary Figure 6) with respect to reference measurements. The distribution of composition values is shifted towards leaner Zn and Mg levels, which are reduced to as low as 4.7 at. % and 10.6 at. % respectively, whereas Cu increases to as much as 18.9 at. %. However, at 4, 6 and 8μm ahead of the crack composition levels fall within the reference measurements.

The matrix composition is also altered near the GB, as shown in Figure 6c. Matrix solute levels are slightly elevated 2μm ahead of the crack tip, but close to the reference values reported in Figure 1b. The matrix composition 4μm ahead of the crack represents a twofold increase in Zn and Mg, reaching 0.83 at. % and 0.58 at. % respectively. It should be noted that matrix η-phase precipitates showed unaltered composition values at this very position. Further ahead, at 6μm, the Zn and Mg content in the matrix decreases again but is still moderately elevated, and it drops back to the reference values at 8μm. Cu levels are elevated in the matrix ahead of the crack, regardless of the distance with respect to the crack tip, although fluctuation is also observed. These compositional modifications can be related to the presence of the progressing crack, as they are outside the standard deviation obtained for the reference measurements, indicated by shaded areas surrounding a horizontal dotted line in Figure 6c.

Interestingly, an η-phase GBP in the dataset 4 μm ahead of the crack was found to contain 37 at. % Mg, higher than the 33 at. % equilibrium composition of the η-phase (Mg(Zn,

Cu, Al)$_2$), as shown in Supplementary Figure 7. An η-phase GBP with increased Mg content ahead of a stress-corrosion crack was also recently reported in another 7xxx Al-alloy [16].

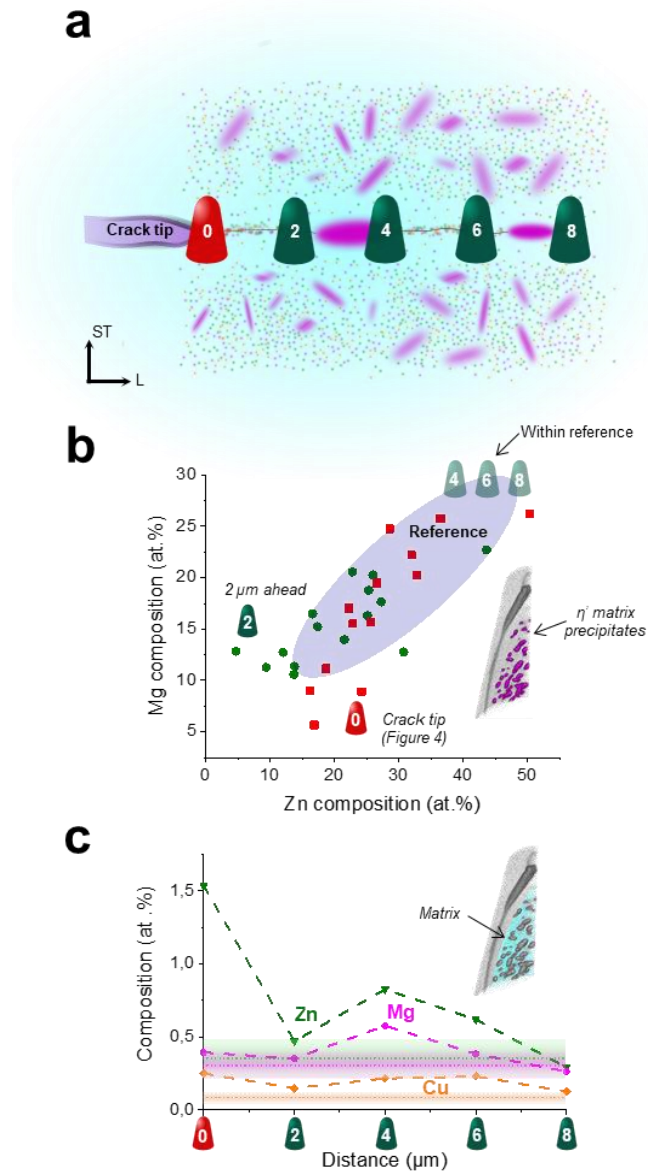

*Figure 6. Evolution of the grain boundary ahead of the stress-corrosion crack. a) Microstructure schematic of a typical 7xxx Al-alloy showing the location of the APT analyses at the crack tip and ahead of the crack. b) Compositional evolution of the matrix η-phase precipitates at the crack tip and ahead of the crack. c) Evolution of the matrix composition with the distance to the crack tip. The dotted lines correspond to the matrix reference measurements shown in Figure 1b, for the T/2 position. The graded zones correspond to the standard deviation, also shown in Figure 1b. Note that all datasets were obtained from a single liftout and that datasets at 2,4 and 6µm contained part of the same grain boundary ahead of the crack.*

## 4 Discussion

Let us first summarize the findings and propose a schematic depiction of some of the phenomena that occur concurrently during SCC and that we could study. Secondly, we will detail aspects of the mechanisms, and provide an interpretation based on our results, keeping in mind that the complexity of the processes at play makes it nearly impossible to provide a holistic perspective of SCC. Throughout the discussion, we will refer to the microstructural features presented in Figure 1.

## 4.1 Overview

Our interpretation of the compositional alterations both at and ahead of the crack tip is schematized in Figure 7 for both T/2 and T/4 in the bottom and top part respectively. The schematic and the following explanation are structured based on the order of microstructural processes as the stress-corrosion crack progresses.

When the crack is several µm away, the microstructure is in the near-equilibrium state that results from the ageing heat treatment (Figure 7a). Then, dislocation activity intensifies as the stress-corrosion crack advances, increasing precipitate-dislocation interactions. Some of these matrix η-phase precipitates can be sheared, which may contribute to destabilizing them and favor their dissolution back into the matrix. This is generally referred to as strain-induced precipitate dissolution and is a common microstructural observation after e.g. fatigue [33]. The matrix and η-phase precipitates' composition ahead of the crack would hence be modified (Figure 6). Simultaneously, solutes at the GB diffuse towards the crack tip, following the gradient of chemical potential established by the preferential oxidation of Mg. This drives the solutes released into the matrix ahead of the crack towards the GB, possibly assisted by non-equilibrium segregation facilitated by the vacancies injected during plastic deformation [34] or via the network of dislocations acting as accelerated diffusion paths, i.e. pipes [35]. This can feed into the flux of solutes towards the crack tip, as the GB is replenished from the adjacent matrix, itself replenished by the dissolution of matrix η-phase precipitates. The η-phase GBPs are also undergoing compositional modifications, with the Mg content increasing to above the equilibrium value (Supplementary Figure 7). This hypothesized sequence is depicted in Figure 7b. Figure 7c shows the evolution after progression of the crack tip and oxidation on the freshly exposed surface starts, triggering further dissolution of matrix η-phase precipitates. The microstructure now presents compositional alterations both within the grain (Figure 6) and at the GB (Supplementary Figure 7). The solutes released by the matrix η-phase precipitates subsequently diffuse towards the growing oxide, thereby enriching the matrix in Zn and Cu as Mg is preferentially oxidized (Figure 4), thereby altering its electrochemical behavior. This will be referred to as oxidation-induced precipitate dissolution, and was recently reported to be occurring during SCC of a 7xxx Al-alloy [16].

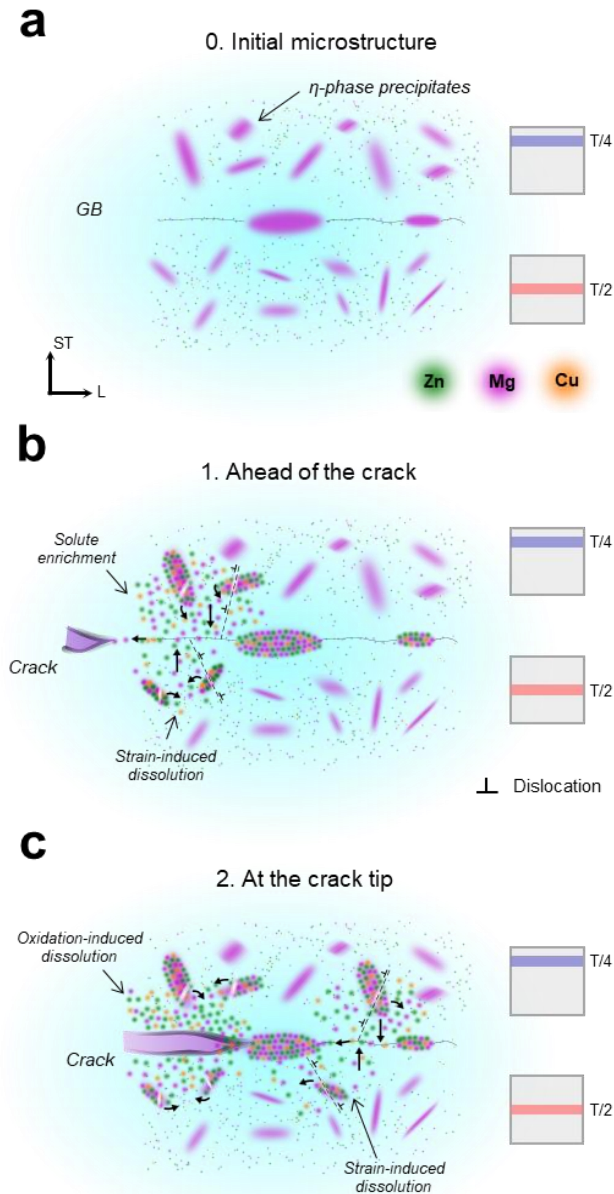

*Figure 7. Mechanistic interpretation of the evidence at both plate positions. a) state of the microstructure after the ageing heat treatment, with the sample from T/4 displaying a higher matrix η-phase precipitate volume fraction. b) schematic description of strain-induced dissolution of matrix η-phase precipitates ahead of the crack and its subsequent segregation to the GB and diffusion towards the crack tip. c) comprehensive schematic description of the events at the crack tip. Matrix η-phase precipitates are further dissolved through oxidation-induced precipitate dissolution and the matrix adjacent to the oxide is enriching in solute elements. At the GB, η-phase GBPs oxidize while, concurrently, solutes released ahead of the crack through strain-induced dissolution are diffusing towards the crack tip.*

### 4.2 Ahead of the crack

APT analyses of the GB located 2–8 µm ahead of the crack tip (Figure 6a) revealed that the microstructure was modified with respect to reference measurements. The solute content in the matrix was elevated and the matrix η-phase precipitates were leaner than expected (Figure 6b–c), and hence likely less thermodynamically stable than their equilibrium composition. Although other mechanistic explanations are plausible, we propose that strain-induced precipitate dissolution [33] contributes to the partial dissolution of matrix η-phase precipitates. Dislocation activity ahead of the crack can shear precipitates, increasing their free energy through the creation of an antiphase boundary and interfacial steps [36]. Although this has not been directly imaged here, repeated shearing often leads to dissolution, as is frequently observed within persistent slip bands in precipitation strengthened alloys after cyclic deformation [36–40]. Strain also produces excess vacancies ahead of the crack [41], with recent studies indicating a five order of magnitude increase in diffusivity in Al strained at $10^{-2}$ $s^{-1}$ and at 100°C [42]. These additional vacancies can then favor non-equilibrium GB segregation of the newly released solutes [34], thereby increasing solute content at the GB. Pipe diffusion might also contribute to transport between the matrix and the GB or within the matrix [35,43,44]. Strain-induced precipitate dissolution ahead of the crack would be more pronounced in higher-strength 7xxx Al-alloys, as they present an increased matrix precipitate number density that confers them with increased mechanical properties.

Overall, the evidence presented in Figure 6 and discussed above suggests that, ahead of the crack, there is a solute influx from the matrix η-phase precipitates towards the GB which subsequently diffuse to the crack tip also along the GB. This is further supported by the observed composition gradient ahead of the crack and shown in Supplementary Figure 8a–b. These are measurements of the total solute composition within the grain interiors, including matrix and precipitates. Solute levels reach a minimum 2μm ahead of the crack, and progressively increase further ahead. These fluxes of solutes from the grain interiors could compensate the loss of solutes at the GB associated to the flux towards the crack, driven by the preferential oxidation of Mg near the crack tip triggering diffusion. Interestingly, the distribution for Zn and Cu is similar to that of Mg. The Ellingham diagram [45] suggests that these species are not subject to preferential oxidation compared to Al, even though Zn is incorporated into the oxide (Figure 4). The measured Zn/Cu distribution could be related to e.g. kinetic effects or stress-assisted diffusion towards the stressed region at the crack tip, as reported to occur in a Fe/Zn system and with H in steel [46,47]. Matrix diffusion does not seem to be the preferred diffusion pathway, as one would measure an opposite composition gradient to what is shown in Supplementary Figure 8b, with a higher total solute content closer to the crack tip. Further, solute levels at the GB, plotted in Supplementary Figure 8c, present near constant Cu levels, but Mg and notably Zn present fluctuations, as Zn reaches 1 at. % 6μm ahead of the crack but is reduced to 0.7 at. % at the 2 and 4μm positions. These measurements were based on the same datasets as those used in Figure 6.

### 4.3 At the crack tip

Analyses of the crack tips at the two plate positions displayed consistent, reproducible observations (Figure 4). We report on the formation of a Zn- and Mg- rich hydroxide at the crack tip (Figure 4b-c). The matrix adjacent to the oxide is enriched in solute (Figure 5) whereas the nearby matrix η-phase precipitates are leaner in Zn, Mg and Cu (Figure 6b and Supplementary Figure 6). There are also qualitative similarities with measurements performed in another 7xxx Al alloy after SCC in Cl-containing solution [16], suggesting that similar mechanisms are at play. This observation is surprising, as the severity of oxidation/corrosion in NaCl is higher than in hot humid conditions. Upon crack advancement, a fresh surface is exposed and oxidation of the GB, GBPs and nearby matrix commences, triggering the dissolution of nearby matrix η-phase precipitates. A detailed explanation of the possible contribution of oxidation to precipitate dissolution can be found elsewhere [16]. Care should be taken when evaluating the SCC susceptibility of different alloys based on near-equilibrium compositions obtained after the ageing heat treatment.

### 4.4 Crack growth behavior

The complex interplay between strain, oxidation, diffusion, grain structure and texture results in a dissimilar SCC crack growth behavior, with the T/4 sample exhibiting a lower $K_{ISCC}$. From a purely mechanical perspective, the flatter, more aligned grains measured at T/4 will favor crack propagation through increasing the effective stress intensity at the crack tip and reducing crack deflection. This is evidenced by the higher aspect ratio and lower recrystallized grains fraction measured at T/4 (Supplementary Table 1). The creation of ligaments behind the crack front, through the higher deflection at T/2, could also be a contributing factor to the lower SCC sensitivity at this position, as demonstrated through 3D studies of SCC cracks in 7xxx Al-alloys [21,48]. On a more speculative note, the sharper textures measured at T/2 (Supplementary Figure 2) will reduce the misorientation of the GBs available for cracking, possibly reducing the number of GBs sensitive to SCC. The more marked texture at T/2 will also cause an increased strain localization [49], thus perhaps increasing crack blunting with respect to T/4. Further, should dislocation emission play a role in SCC crack growth of 7xxx Al-alloys, through e.g. the AIDE mechanism [13], the strain localization directions imposed by the texture at T/2 might not lead to crack advancement, thus also contributing to the higher $K_{ISCC}$ measured for T/2.

The characterization of the crack tip microchemistry at both plate positions revealed marked differences, with a more pronounced crack tip matrix solute enrichment at the T/4 position (Figure 5). This is presumably caused by the higher total solute content at T/4 (Table 2), which upon strain- and oxidation-induced dissolution of the matrix η-phase precipitates and subsequent diffusion processes would end up near the crack tip. Solutes in solid solution in Al can change its electrochemical properties in Cl-containing solutions, especially the dissolution kinetics and the corrosion and pitting potentials [50–54]. Zn increases the dissolution kinetics while Cu lowers them and Mg additions don't have any notable effect. Although corrosion studies performed in Cl-containing solutions may not be directly translatable to oxidation in hot humid air, these can be used to hypothesize. The higher Zn content in solid solution near the crack tip at T/4 (Figure 5) could increase oxidation rates and H production at the crack tip with respect to T/2. Zn incorporation in the oxide would decrease its stability thus rendering is hydration easier, as proposed by Song et al. [55] to explain the activation of Al by Zn, although this was not directly investigated experimentally. A higher concentration of atomic H within the region ahead of the crack might then facilitate crack growth, thus requiring less mechanical driving force to achieve crack advancement. Further, a faster growing oxide could accelerate the increase in local stress intensity — the "wedge effect" — due to the higher crack opening displacement this would induce. The lower $K_{ISCC}$ measured for the T/4 DCB sample (Figure 2) could therefore be explained through a synergistic effect of both a faster "wedge effect" and an increased H generation.

Conversely, at the higher stress intensities characteristic of region II, the crack growth rate is nearly identical at both plate positions (Figure 2), suggesting that there is another process rather than oxidation rates or mechanical effects that is limiting the advancement of the crack. It has been previously proposed [48], that the plateau observed in region II is caused by a balance of the driving forces for crack propagation with the metal ligaments and the interactions due to the multiplicity of cracks. But we report on nearly identical crack growth rates for the two plate positions, which, at the same time, exhibit grain structures with different propensities to crack branching and therefore, to the creation of ligaments. This can be due to differences in microbranching present at low stress intensities [56] or to mechanisms that remain unidentified. Holroyd and Scamans [4] previously suggested that another, strain-related process could also be rate-controlling.

Another alternative explanation is that the microchemical environment at the crack tip, and possibly the amount of H generated, is relatively similar in both samples in region II loading conditions, as the microstructure evolution ahead of the crack and at the crack tip presented herein would be caused by time-dependent processes. Knight et al. in 2010 [57], measured a correlation between region II crack growth rates and the electrochemical potential of the GB, with more active GBs resulting in faster crack velocities. As crack growth rates in region II were identical for both plate positions in the present study, it is reasonable to assume that the GB at T/2 and at T/4 would have similar electrochemical activity, despite the slight differences in composition (Figure 1). By definition, the lower crack growth rates of region I will give more time in-between crack advance events for the microstructure-altering processes proposed herein to occur. This would then presumably allow the subtle composition differences between plate positions (Table 2 and Figure 1) to have a higher effect on H generation and oxidation rates at the crack tip at lower stress intensity values.

We could not directly evaluate the role of the GB and its precipitates on the cracking performance mainly because of the lack of statistically representative composition measurements. Although we measured compositional differences in the η-phase GB precipitates between the two plate positions, its effect on the electrochemical properties of the GB as a whole is unclear. To properly assess this a characterization of the GB equivalent to the thorough investigation by Garner et al. [58] is needed, along with electrochemical measurements of the GBs [57].

The electrochemical behavior of the GB, and of η-phase GBPs in particular, is expected to control H production at stress-corrosion crack tips. In humid air, η-phase GBPs or particles like $Mg_2Si$ can self-oxidize or establish short-range galvanic interactions with the matrix [32]. Interestingly, we report solute enrichment within the matrix near the crack tip as a consequence of the oxidation-induced dissolution of matrix η-phase precipitates (Figure 4). Therefore, the matrix is oxidizing independently of the η-phase GBPs, highlighting the importance of understanding the effect of alloying elements in an Al solid solution upon the oxidation behavior. If we consider that the near GB matrix would represent a substantial amount of the oxidizing surface at the crack tip [58], its contributing role to oxidation at the crack tip cannot be overlooked. In addition, based solely on the fact that matrix η-phase precipitates are releasing solutes both ahead (Figure 6 and Supplementary Figure 8) and at the crack tip (Figure 5), we can suggest that its composition, number density and volume fraction may all be relevant.

Based on the arguments outlined above, if the matrix microstructure and microchemistry have an effect on SCC behavior, it would be expected at lower crack growth rates, where more time is available between crack advances. This is consistent with the crack growth rate measurements presented above (Figure 2). However, it is unknown how much of a role these would play and the near GB microchemistry is still expected to be the primary determining factor for H generation and oxidation rates. Differences in texture, grain structure, crack branching and oxide growth rates could affect the effective stress intensity at the crack tip, thereby altering the mechanical driving force for crack propagation.

We also hypothesized that differences in the active HE mechanisms between plate positions could explain the contrasting SCC behavior observed in region I. But we could not rationalize this based solely upon the current available literature on HE mechanisms. Therefore, we believe that the driving factors behind the observed contrasting SCC performances is a combination of differences in H generation at the crack tip and mechanical effects. Although with the evidence presented herein, we could not establish their relative contributions to crack growth. Nevertheless, many other microstructural parameters and active processes are playing a critical role and its possible implications should not be neglected.

# 5 Conclusions

We studied the influence of grain structure and composition on stress-corrosion cracking of 7xxx Al-alloys in hot humid air, using atom probe tomography and double cantilever beam crack growth tests. Here, we have shown that the microchemical environment at the crack tip presents an enriched matrix solute content and the dissolution of matrix η-phase precipitates. This work is consistent with the similar observations reported in another 7xxx Al-alloy after SCC in Cl solution [16].

We measure substantial differences in $K_{ISCC}$ but identical crack growth rates in region II between these samples extracted from T/2 and T/4 and subsequently processed to a non-industrially relevant T6 temper. We conclude that the observed dissimilar behavior in region I may be due to differences in oxidation rates at the crack tip and to mechanical effects, related to the grain structure at each plate position. We have also revealed composition alterations several μm ahead of the crack, within the η-phase GBPs and also in the matrix η-phase precipitates presumably through strain-induced dissolution. Solutes released ahead of the crack could then diffuse towards the crack tip along the GB, further altering its electrochemical activity.

In summary, we postulate that, in region I loading conditions, a dynamic view of the microstructure at the GB and the nearby grain interior might be crucial to understanding the role of the alloy's chemistry in SCC, especially the role of Zn and Mg. The extent to which chemical and mechanical effects are determining SCC susceptibility is currently not established but it is likely that other known or unknown factors are at play.

# 6 Author contributions

L.P. designed the study and oversaw DCB testing. M.L.F. prepared atom probe specimens, SEM and EBSD samples. M.L.F. performed APT, SEM and EBSD analysis, processed and interpreted the data. L.P., T.W. and B.G. contributed to the interpretation of the data. M.L.F. and B.G. wrote the manuscript. L.P. and T.W. reviewed the manuscript.

# 7 Acknowledgements

We thank U. Tezins, A. Sturm, M. Nellessen, C. Broß, and K. Angenendt for their technical support at the FIB/APT/SEM facilities at MPIE. We are grateful for conducting the DCB testing to Romain Bergeron and Annabelle Rossetto B.G. acknowledge the financial support from the ERC-CoG-SHINE-771602.

# 9 Supplementary information

Supplementary Table 1. Analysis of the EBSD data shown in Supplementary Figure 1

|  | Grain size [µm2] | Aspect ratio (Area fraction) | Recrystallized fraction |
|---|---|---|---|
| T/2 | 83580,7 | 6.7 | 22.3 |
| T/4 | 77005.6 | 8.3 | 20.1 |

Supplementary Table 2. Area fractions of texture components

|  | Copper [% area fraction] | S3 [% area fraction] | Brass [% area fraction] | Goss [% area fraction] |
|---|---|---|---|---|
| T/2 | 3.1 | 13.0 | 11.0 | 1.4 |
| T/4 | 1.8 | 1.9 | 0.5 | 0.1 |

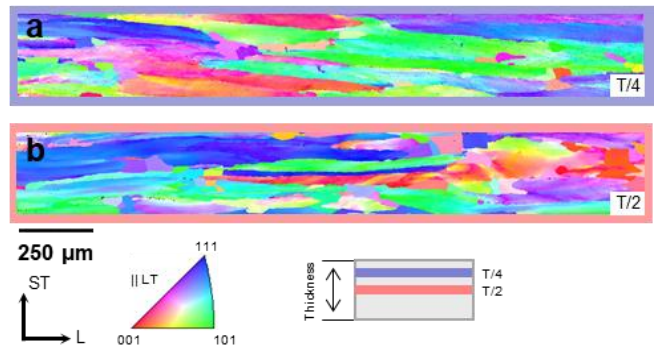

Supplementary Figure 1. Representative regions of the 2.1 x 0.95 mm EBSD scans for a) T/4 and b) T/2.

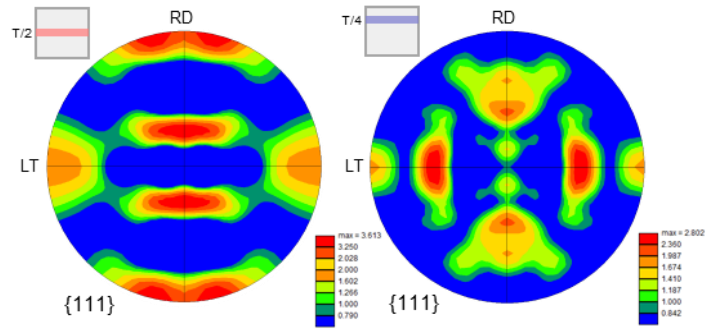

Supplementary Figure 2. {111} pole figures for the two plate positions.

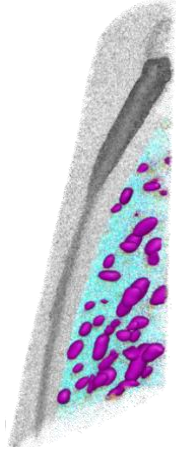

*Supplementary Figure 3. APT reconstruction of a GB containing dataset. The coloured features display the location of the measurements shown in Table 2.*

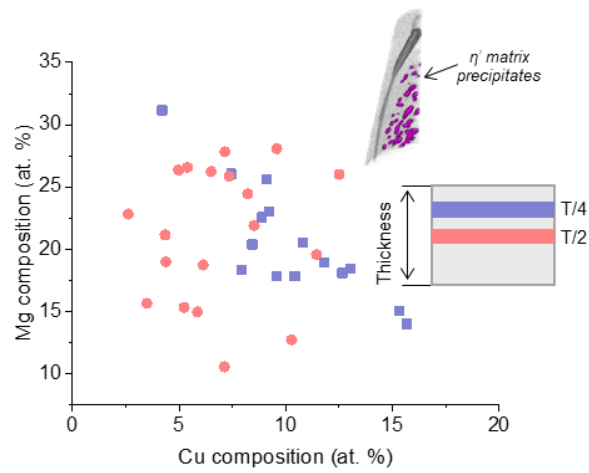

*Supplementary Figure 4. Compositions of the matrix η-phase precipitates within the grain interior at both plate positions. 2 datasets for each plate position are shown and each data point corresponds to an individual matrix η-phase precipitate.*

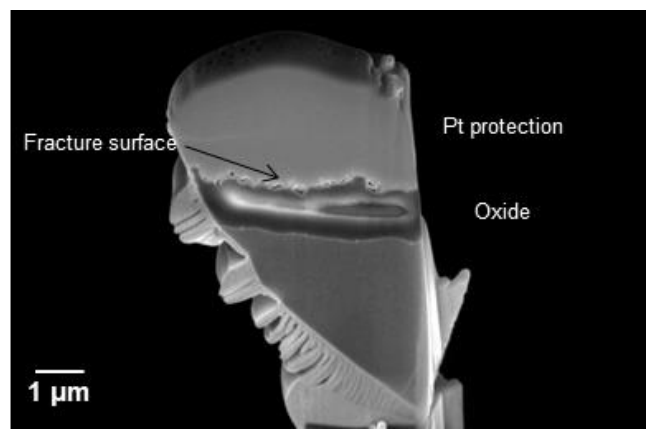

*Supplementary Figure 5. PFIB liftout from the fracture surface showing the thickness of the oxide layer.*

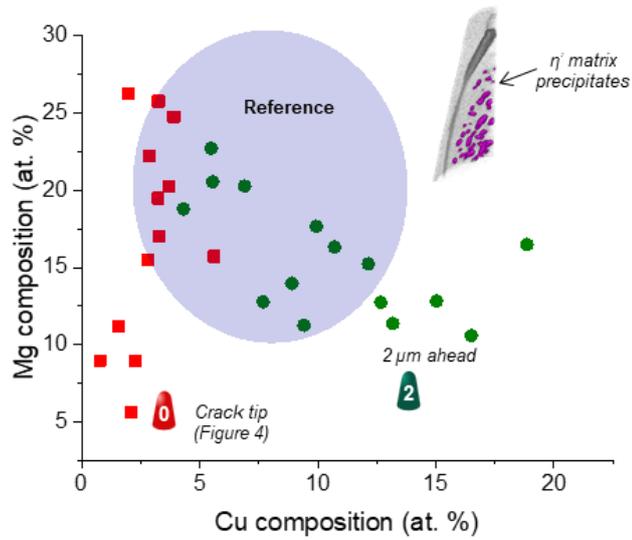

*Supplementary Figure 6. Compositional evolution of the matrix η-phase precipitates at the crack tip and ahead of the crack*

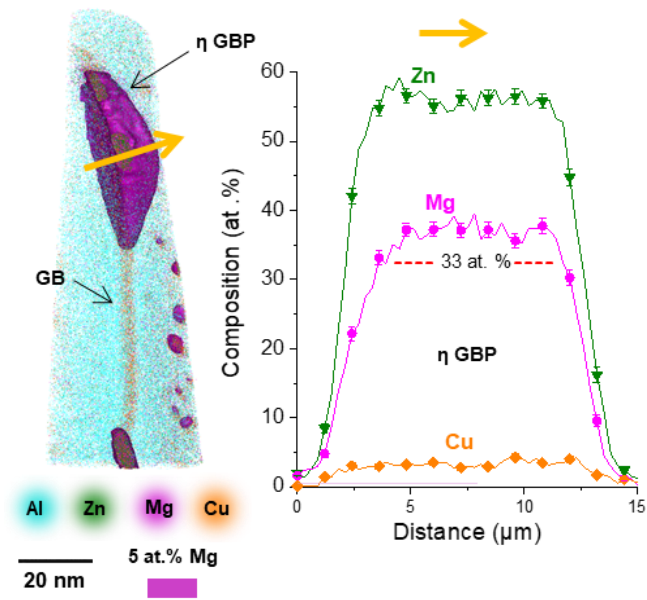

*Supplementary Figure 7. APT reconstruction from 4μm ahead of the crack showing an η-phase GBP with an elevated Mg composition.*

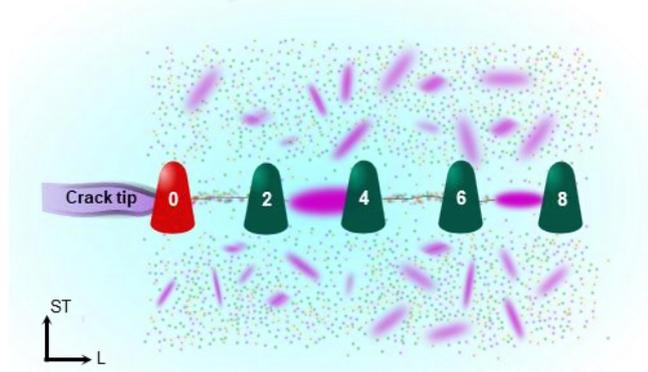
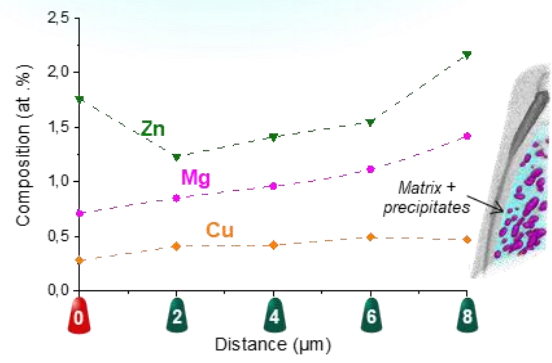
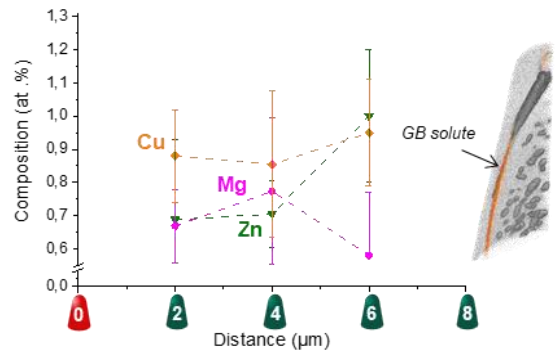

*Supplementary Figure 8*